\documentclass[11pt]{article}
\def\be{\begin{equation}}
\def\ee{\end{equation}}
\def\ba{\begin{eqnarray}}
\def\ea{\end{eqnarray}}
\def\la{\langle}
\def\ra{\rangle}
\def\a{\alpha}

\def\h{\hskip 1cm}

\def\lo{\longrightarrow}
\def\A1{A_{-1}}
\usepackage{epsfig}
\begin{document}
\begin{titlepage}

\begin{center}{\Large \bf A classification of spin 1/2 matrix product states with two dimensional auxiliary matrices}\\
\vspace{2cm}\h Marzieh Asoudeh
\footnote{$m-asoudeh@sbu.ac.ir$}\vspace{1cm} $^{\dag}$ \\
Department of Physics, Shahid Beheshti
University, GC\\
19839-63113, Tehran, Iran\\
\end{center}

\vskip 2cm

\begin{abstract}
We classify the matrix product states having only spin-flip and
parity symmetries, which can be constructed from two dimensional
auxiliary matrices. We show that there are three distinct classes of
such states and in each case, we determine the parent Hamiltonian
and the points of possible quantum phase transitions. For two of the
models, the interactions are three-body and for one the interaction
is two-body.
\end{abstract}
\vskip 2cm PACS Number:75.10.Jm \hspace{.3in}
\end{titlepage}

\section{Introduction}\label{intro}
The problem of determining the ground state of a given many-body
Hamiltonian, is an important problem in condensed matter and
mathematical physics. There is already a rich literature on this
subject, which dates back to the work of Hans Bethe  on the
Heisenberg spin chain and has continued since then with the works of
many other people including Yang, Baxter, and Lieb, to name only a
few. In particular for spin systems, the exponential increase in the
dimension of Hilbert space of such a system, as the number of
particles rise, turns this problem into a computationally formidable
one, beyond the capability of any classical computer. In fact it has
now been established that finding the ground state of a given
many-body Hamiltonian is the analog of NP-complete problems for
quantum computers \cite{Nielsen}. The lesson that we learn from all
this is that it is highly improbable that we be able to find a
generic system with exactly known ground state. Nevertheless, there
are systems with exactly known ground states and even if such
systems are not exactly what we have in nature or in the
laboratories, they may be good approximations to real systems, or at
least may teach us useful and important concepts and methods for
studying more
realistic systems.\\

One of the methods, developed in recent years, for investigating
this problem is the Matrix Product State or Finitely Correlated
State \cite{fcs1,fcs2,fcs3} formalism. It is also called Optimal
Ground State formalism in some references \cite{ops1,ops2,ops3}. The
main theme is that one starts with a state with prescribed
symmetries and properties, and then construct the family of
Hamiltonians for which this state is an exact ground state. It is
obvious that for any given state $|\psi\ra$, the equation
$H|\psi\ra=0$ has always many solutions for the unknown $H$, since
the number of equations is much less than the number of unknowns.
However the problem becomes interesting and quite non-trivial when
we put physical constraints on the Hamiltonian. That is we demand
that i) $H$ be positive, so that $|\Psi\ra$ is actually the ground
state and not an ordinary eigenstate, ii) that it be a sum of local
terms, i.e. $H=\sum_k h_{k,k+l}$, where $h_{k,k+l}$ acts on a block
of $l+1$ spins, and iii) that both the state and the Hamiltonian
have some reasonable physical symmetries, like parity, spin-flip,
and at times
rotational symmetries. \\

In the past few years, a lot of interest has been attracted to the
subject of matrix product states
\cite{mps1,mps2,mps3,mps4,mps5,mps6}, specially after the emergence
of the field of quantum information \cite{qps1,qps2,qps3,qps4,qps5}.
The reason is the complementary role that the fields of condensed
matter physics and quantum information play in investigation of many
body systems. On the one hand quantum information starts with
properties of states, while condensed matter physics, starts from
the properties of the Hamiltonian which embodies the interactions
and energy of the system. The matrix product formalism is one of the
subjects which
lies at the borderline of these two subjects. \\

As is well known, in this formalism, one starts from proposed states
 whose expansion coefficients are the trace of product of given
matrices. While for numerical investigations, i.e. the density
matrix renormalization group (DMRG), one usually starts from large
dimensional matrices, to simulate ground states of given
Hamiltonians, in the approach which is used for finding exactly
solvable models, one starts from low dimensional matrices and finds
Hamiltonians for which these states are exact ground states. This is
the approach which has been used in our works and in many other
works in the past few years
\cite{mps1,mps2,mps3,mps4,mps5,mps6,mps11,mps12,mps13,mps14,mps15,mps16,mps17}.
In this article we follow this approach and classify all the matrix
product states which can be constructed from two dimensional
matrices. We restrict ourselves to states which allow one or another
of the spin-flip or parity symmetries and find that there are three
classes of such matrix product states. We will study these states
and find the parent Hamiltonians and also the points or lines in the
space of control parameters where a MPS-quantum phase transition
\cite{Sachdev} (MPS-QPT) may occur. This is a term, introduced in
\cite{Cirac} to differentiate these kinds of QPT's (characterized by
any discontinuity in any macroscopic quantity) from the conventional
QPT,s in which a non-analyticity in the ground state energy
typically occurs.  \\

The structure of this paper is as follows: To make the article
self-contained, in the next section we briefly introduce the basic
elements of the formalism.  In section (\ref{classify}) we discuss
the symmetry properties of MPS and classify the spin 1/2 states with
two dimensional matrices. We show that there are three classes,
denoted by a model A, model B and model C and studied in subsequent
sections. We end the paper with a discussion.

\section{A brief introduction to matrix product states}\label{MPS}
First let us review the basics of matrix product states. Consider a
homogeneous ring of $N$ sites, where each site describes a $d-$level
state. The Hilbert space of each site is spanned by the basis
vectors $|i\ra, \ \ i=0,\cdots d-1$. A state
\begin{equation}\label{state}
    |\Psi\ra=\sum_{i_1,i_2,\cdots i_N}\psi_{i_1i_2\cdots
    i_N}|i_1,i_2,\cdots, i_N\ra
\end{equation}
is called a matrix product state if there exists $D$ dimensional
complex matrices  $A_i\in {C}^{D\times D},\ \ i=0\cdots d-1$ such
that
\begin{equation}\label{mat}
    \psi_{i_1,i_2,\cdots
    i_N}=\frac{1}{\sqrt{Z}}tr(A_{i_1}A_{i_2}\cdots A_{i_N}),
\end{equation}
where $Z$ is a normalization constant given by
\begin{equation}\label{z}
    Z=tr(E^N)
\end{equation}
and
\begin{equation}\label{E}
E:=\sum_{i=0}^{d-1} A_i^*\otimes A_i.
\end{equation}
Here we are restricting ourselves to translationally invariant
states, by taking the matrices to be site-independent. By defining
the vector valued matrix
\begin{equation}\label{Avector}
    {\cal A}=\sum_{i=1}^d A_i |i\ra,
\end{equation}
one can write the MPS in a more concise way as
\begin{equation}\label{concise}
    |\psi\ra = tr({\cal A}\otimes {\cal A}\otimes \cdots {\cal A}),
\end{equation}
where we use the convention $$tr({\cal A}\otimes {\cal A}):=tr(A_i
A_j)|i\ra\otimes |j\ra.$$

It is important to note that the MPS representation (\ref{mat}) is
not unique and a transformation such as
\begin{equation}\label{guage}
A_i\lo \mu UA_i U^{-1}
\end{equation}
where $U$ is an invertible matrix, and $\mu$ is a constant, leaves
the state invariant. The simple structure of the MPS allows also an
easy calculation of correlation functions. Let $O$ be any local
operator acting on a single site. Then we can obtain the one-point
function on site $k$ of the chain $\la \Psi|O(k)|\Psi\ra $ as
follows:
\begin{equation}\label{1point}
    \la \Psi|O(k)|\Psi\ra = \frac{tr(E^{k-1}E_O E^{N-k})}{tr(E^N)},
\end{equation}
where
\begin{equation}\label{mpsop}
E_O:=\sum_{i,j=0}^{d-1}\la i|O|j\ra A_i^*\otimes A_j.
\end{equation}
In the thermodynamic limit $N\lo \infty$, equation (\ref{1point})
gives
\begin{equation}\label{1pointthermo}
    \la \Psi|O|\Psi\ra = \frac{\la
    \lambda_{max}|E_O|\lambda_{max}\ra}{\lambda_{max}},
\end{equation}
where we have used the translation invariance of the model and
$\lambda_{max}$ is the eigenvalue of $E$ with the largest absolute
value and $|\lambda_{max}\ra$ and $\la \lambda_{max}|$ are the right
and left eigenvectors corresponding to this eigenvalue, normalized
such that $\la \lambda_{max}|\lambda_{max}\ra=1$. Here we are
assuming that the largest eigenvalue of $E$ is non-degenerate. In
case $\lambda_{max}$ is degenerate with degree equal to $g$, then
Eq. (\ref{1pointthermo}) will be modified to
\begin{equation}\label{1pointthermoNew}
    \la \Psi|O|\Psi\ra = \frac{\sum_{i=1}^g\la
    \lambda_{max,i}|E_O|\lambda_{max,i}\ra}{\lambda_{max}},
\end{equation}

The n-point functions can be obtained in a similar way. For example,
the two-point function $\la \Psi|O(k)O(l)|\Psi\ra$ can be obtained
as
\begin{equation}\label{2point}
\la \Psi|O(k)O(l)|\Psi\ra = \frac{tr(E_O(k)E_O(l)E^N)}{tr(E^N)}
\end{equation}
where $E_O(k):=E^{k-1}E_OE^{-k}$. Note that this is a formal
notation which allows us to write the n-point functions in a uniform
way, it does not require that $E$ be an invertible matrix. In the
thermodynamic limit the two point function turns out to be
\begin{equation}\label{2pointThermodynamicLimit}
\la \Psi|O(1)O(r)|\Psi\ra = \frac{1}{\lambda_{max}^{r}} {\sum_i
\lambda_i^{r-2} \la\lambda_{max}|E_{O}|\lambda_{i}\ra\la
\lambda_i|E_{O}|\lambda_{max}\ra}.
\end{equation}
For large distances $r\gg 1$, this formula reduces to
\begin{equation}\label{2pointrLarge}
\la \Psi|O(1)O(r)|\Psi\ra-\la \Psi|O|\Psi\ra^2
=\frac{\lambda_1^{r-2}}{\lambda_{max}^r} {\la
\lambda_{max}|E_{O}|\lambda_{1}\ra\la
\lambda_{1}|E_{O}|\lambda_{max}\ra},
\end{equation}
where $\lambda_1$ is the second largest eigenvalue of $E$ for which
the matrix element $\la \lambda_1|E_O|\lambda_{max}\ra$ is
non-vanishing and we have assumed that the eigenvectors of $E$ have
been normalized, i.e. $\la \lambda_i|\lambda_j\ra = \delta_{ij}$.
Thus the correlation length is given by
\begin{equation}\label{corr}
    \xi = \frac{1}{\ln \frac{\lambda_{max}}{\lambda_{1}}}.
\end{equation}

Any level crossing in the largest eigenvalue of the matrix $E$
signals a possible MPS-QPT . Here we are using the term quantum
phase transition in a broader sense than usual \cite{Cirac}, that
is, we call any discontinuity in any macroscopic quantity a quantum
phase transition, even if the ground state energy itself is a
continuous function of the coupling constants.  Also, due to
(\ref{corr}), any level crossing in the second largest eigenvalue of
$E$ implies the correlation length of the system has undergone a
discontinuous change.

\subsection{The Hamiltonian}
Given a matrix product state, the reduced density matrix of $k$
consecutive sites is given by
\begin{equation}\label{rhok}
    \rho_{i_1\cdots i_k, j_1\cdots j_k}=\frac{tr((A_{i_1}^*\cdots A_{i_k}^*\otimes A_{j_1}\cdots A_{j_k})E^{N-k})}{tr(E^N)}.
\end{equation}
The null space of this reduced density matrix includes the solutions
of the following system of equations
\begin{equation}\label{cc}
    \sum_{j_1,\cdots, j_k=0}^{d-1}c_{j_1\cdots
    j_k}A_{j_1}\cdots A_{j_k}=0.
\end{equation}
Given that the matrices $A_i$ are of size $D\times D$, there are
$D^2$ equations with $d^k$ unknowns. Since there can be at most
$D^2$ independent equations, there are at least $d^k-D^2$ solutions
for this system of equations. Thus for the density matrix of $k$
sites to have a null space it is sufficient that the following
inequality holds
\begin{equation}\label{dD}
    d^k\ >\ D^2.
\end{equation}
Let the null space of the reduced density matrix be spanned by the
orthogonal vectors $|e_{\a}\ra, \ \ \ (\a=1, \cdots  s,\geq
d^k-D^2)$. Then we can construct the local hamiltonian acting on $k$
consecutive sites as
\begin{equation}\label{h}
    h:=\sum_{\a=1}^s \mu_{\a} |e_{\a}\ra\la e_{\a}|,
\end{equation}j
where $\mu_{\a}$'s are positive constants. These parameters together
with the parameters of the vectors $|e_i\ra $ inherited from those
of the original matrices $A_i$, determine the total number of
coupling constants of the Hamiltonian.  If we call the embedding of
this local Hamiltonian into the sites $l$ to $l+k$ by $h_{l,l+k}$
then the full Hamiltonian on the chain is written as
\begin{equation}\label{H}
    H=\sum_{l=1}^N h_{l,l+k}.
\end{equation}
The state $|\Psi\ra$ is then a ground state of this hamiltonian with
vanishing energy. The reason is as follows:
\begin{equation}\label{Hrho}
\la \Psi|H|\Psi\ra=tr(H|\Psi\ra\la\Psi|)=\sum_{l=1}^N
tr(h_{l,l+k}\rho_{l,l+k})=0,
\end{equation}
where $\rho_{l,k+l}$ is the reduced density matrix of sites $l$ to
$l+k$ and in the last line we have used the fact that $h$ is
constructed from the null eigenvectors of $\rho$ for $k$ consecutive
sites. Given that $H$ is a positive operator, this proves the
assertion.\\

In view of the above introduction, we have a clear recipe for
constructing matrix product states and a family of parent
Hamiltonians. First one chooses the matrices throwing away all
spurious degrees of freedom by transformations (\ref{guage}) and
reducing further the degrees of freedom by imposing symmetries. In
this way one ends with a reasonable set of matrix product states,
which hopefully may have applications in description of real
physical systems. Imposing a continuous symmetry, like rotation
around an axis, restricts the matrices considerably
\cite{mps1,mps2,mps3}. In this article we restrict ourselves to
discrete symmetries only which allow a larger variety of models to
be constructed. For two dimensional auxiliary matrices, this is a
simple tractable problem, which we do in this article. For larger
matrices, the problem is not so simple and we defer it to another
work.

\section{The classification of 2 dimensional matrices for
matrix product states of spin 1/2 chains }\label{classify}

We now classify all the two dimensional matrices which can be used
for constructing spin 1/2 matrix product states. We restrict
ourselves to the case where these states have spin-flip and
left-right symmetries.

\subsection{Symmetries}
Consider now a local symmetry operator $R$ acting on a site as
$R|i\ra=R_{ji}|j\ra$ where summation convention is being used. $R$
is a $d$ dimensional unitary representation of the symmetry. A
global symmetry operator ${\cal R}:=R^{\otimes N}$ will then change
this state to another matrix product state
\begin{equation}\label{mpsPrime}
    \Psi_{i_1i_2\cdots i_N}\lo \Psi':=tr(A'_{i_1}A'_{i_2}\cdots
    A'_{i_N}),
\end{equation}
where
\begin{equation}\label{A'}
    A'_i:=R_{ij}A_j.
\end{equation}
A sufficient but not necessary condition for the state $|\Psi\ra$ to
be invariant under this symmetry is that there exist an operator
$U(R)$ such that
\begin{equation}\label{symm}
    R_{ij}A_j=U(R)A_iU^{-1}(R).
\end{equation}
Thus $R$ and $U(R)$ are two unitary representations of the symmetry,
respectively of dimensions $d$ and $D$.  Equation (\ref{symm}) will
be our guiding lines in defining states with prescribed symmetries.
Spin-flip symmetry means that
\begin{equation}\label{parity}
    \psi_{\overline{i_1},\overline{i_2},\cdots \overline{i_N}}=\psi_{i_N,i_{N-1},\cdots i_1},
\end{equation}
where $\overline{i}=1-i.$
For a matrix product state, this requires
that there be a matrix like $X$, such that
\begin{equation}\label{X}
    XA_0X^{-1}=\epsilon A_1, \h XA_1X^{-1}=\epsilon A_0,
\end{equation}
where $\epsilon=\pm 1$. Similarly left-right symmetry means that

\begin{equation}\label{parity}
    \psi_{i_1,i_2,\cdots i_N}=\psi_{i_N,i_{N-1},\cdots i_1}.
\end{equation}

For a matrix product state, this means that there be a matrix
$\Omega$ such that
\begin{equation}\label{Omega}
    \Omega A_0\Omega^{-1}=\sigma A^{T}_0, \h \Omega A_1\Omega^{-1}=\sigma
    A^T_1,
\end{equation}
where the superscript $T$ stands for the transpose and $\sigma=\pm
1$. These conditions are general irrespective of the dimension of
matrices. For two dimensional matrices however, if we take the trace
and determinants of both sides of equations (\ref{X}) and
(\ref{Omega}), and comparing them, we find that
\begin{equation}\label{trace}
    tr(A_0)=\epsilon tr(A_1),
\end{equation}
and
\begin{equation}\label{det}
    \det(A_0)=\det(A_1).
\end{equation}
The important point is that for two dimensional matrices, the trace
and determinant are the only invariants under similarity
transformations, and hence these two equations allow us to classify
all the matrices $A_0$ and $A_1$ which can be used for construction
of spin 1/2 matrix product states. We will use the freedom
(\ref{guage}) and also the above two conditions to show that there
are three distinct classes of matrix pairs and corresponding matrix
product states. In the next three sections, we introduce the matrix
pairs and study the properties of the matrices obtained from them.
For ease of distinction, we use different notations for the matrix
pairs in each section, namely we denote the matrix pairs by $A_i$,
$B_i$ and $C_i$ in the following sections.

\section{Model A}\label{ModelA}
If one of the matrices say $A_0$ is diagonalizable, we can use
freedom in re-scaling $A_0\lo \mu A_0$ to put it in the form
\begin{equation}\label{matrix0}
A_{0}=\left(\begin{array}{cc} 1+g & 0 \\ 0 & 1-g \end{array}\right),
\end{equation}
where $g$ is a free parameter. Now take $A_1$ as an arbitrary matrix
of the form $A_1 = \left(\begin{array}{cc} a & b
\\ c & d
\end{array}\right)$. If $bc\ne 0$, then  we can use the transformation (\ref{guage}) with
$U=\left(\begin{array}{cc} \sqrt{c} &  \\  & \sqrt{b}
\end{array} \right)$ and a further re-definition of $\sqrt{bc}\lo b$
to put it in the form
\begin{equation}\label{matrix1}
A_{1}=\left(\begin{array}{cc} a & b \\ b & d \end{array}\right).
\end{equation}
For this type of matrices, the MPS is automatically left-right
symmetric, with $\Omega=I$. From equations (\ref{trace}) and
(\ref{det}), we find the following constraints on the parameters,
\begin{equation}\label{h}
    1-g^2=(ad-b^2) \h  2\epsilon =(a+d).
\end{equation}
To solve the second equation, we put
\begin{equation}\label{u}
    a=\epsilon + u, \h b = \epsilon - u,
\end{equation}
which turns the first equation into
\begin{equation}\label{hh}
   g^2=b^2 + u^2,
\end{equation}
which can be solved by the parametrization $u=g\cos \theta$ and
$b=g\sin \theta$. Therefore the final form of the matrices will be
as follows:
\begin{equation}\label{TzSolution}
 A_0=\left(\begin{array}{cc} 1+g &
0 \\ 0 & 1-g \end{array}\right), \ \ \ \ A_1=\left(\begin{array}{cc}
\epsilon+g\cos\theta & g\sin\theta \\ g\sin\theta & \epsilon-
g\cos\theta
\end{array}\right).
\end{equation}

The matrices satisfy the symmetry constraints (\ref{X}) and
(\ref{Omega}) with $\Omega=I$ and

\begin{equation}\label{XA}
    X=\left(\begin{array}{cc} \epsilon \sin \theta & -\epsilon \sin \theta \\ 1-\epsilon \cos \theta & 1+\epsilon \cos
    \theta\end{array}\right).
\end{equation}

We now restrict ourselves to the case $\epsilon=1$ (We do this also
for other models B and C). To study the properties of the
corresponding MPS, we should determine the eigenvalues of the
transition matrix $E_A:=A_0\otimes A_0+A_1\otimes A_1$. For general
values of the parameters, the analytical form of these eigenvalues
are complicated. However we can gain insight by looking at them for
for generic values of the parameter $g$. Figure (\ref{AFigure})
shows the eigenvalues as a function of $\theta$ for two values of
the parameter $g$. The same pattern repeats for other values of $g$.
We first see that there is no level-crossing in the largest
eigenvalue and hence no MPS quantum phase transition in this model.
However for every value of $g$, two points are important. The point
$\theta=\Pi$ (or $-\Pi$), is the point where the largest and the
next to largest eigenvalues become equal. This is a point where
according to (\ref{corr}), the correlation length becomes infinite.
It is seen from (\ref{TzSolution}) that at this point, the matrices
become diagonal $A_0 = diagonal (1+g, 1-g)$ and $A_1 =diagonal
(1-g,1+g)$ and according to (\ref{concise}), the un-normalized state
becomes the sum of two product states, namely
\begin{equation}\label{prod}
    |\psi\ra=|\phi_+\ra^{\otimes N}+|\phi_-\ra^{\otimes N},
\end{equation}
where
\begin{equation}\label{phi}
    |\phi_+\ra=(1+g)|0\ra+(1-g)|1\ra,\h |\phi_-\ra=(1-g)|0\ra+(1+g)|1\ra.
\end{equation}

\begin{figure}
\centering
    \includegraphics[width=11cm,height=9cm,angle=0]{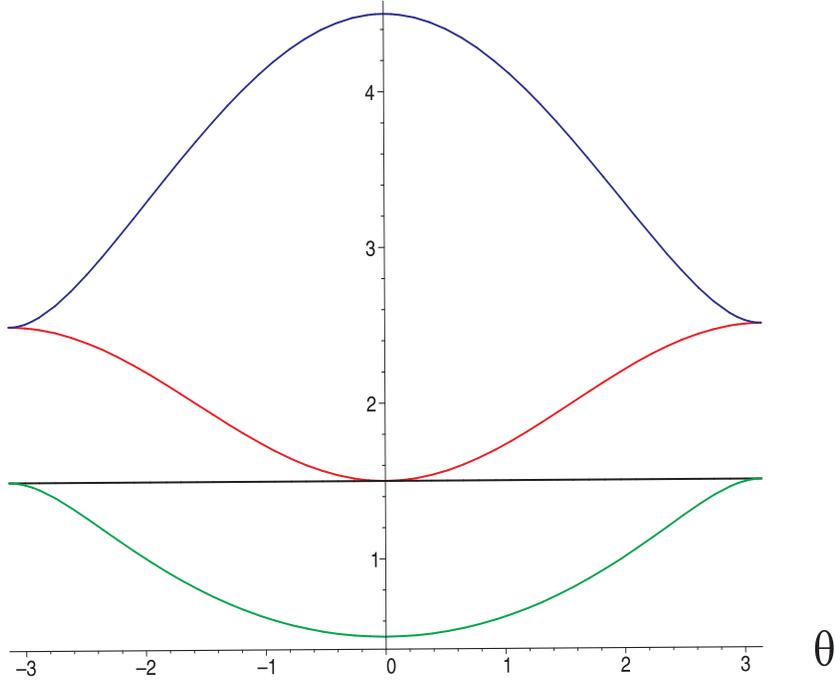}
    \caption{(Color Online) The eigenvalues of the transition matrix $E_A$ for $\epsilon=1$ and a generic value of $g$ (i.e. g=1/2) as a function of $\theta$.
     }
\end{figure}\label{AFigure}

Let us now find the  parent Hamiltonian for a fixed value of the
parameter $g$, say $g=1$. To this end, we have to see for which
value of $k$ (the range of interaction), the system of equations
(\ref{cc}) have a non-trivial solution. It is seen that the smallest
$k$ for which there is such a solution is $k=3$.

The solution space of this system of equations turn out to be
spanned by the following vectors:

\begin{eqnarray}
  |e^{A}_1\ra &=& -\frac{1+\cos(\theta)}{2}|000\ra+|101\ra, \\
  |e^{A}_2\ra &=& |001\ra-|011\ra \\
  |e^{A}_3\ra &=& |100\ra-|110\ra \\
  |e^{A}_4\ra &=& |010\ra-\frac{1+cos(\theta)}{2} |111\ra.
\end{eqnarray}

The symmetries of the parent Hamiltonian now show itself in the form
that the above states, which transform into each other under the
action of these symmetries. To have a Hamiltonian which respects
these symmetries, we construct it as follows:

\begin{equation}\label{hA}
h_A=J(|e^{A}_1\ra\la e^{A}_1|+|e^{A}_4\ra\la
e^{A}_4|)+K(|e^{A}_2\ra\la e^{A}_2|+|e^{A}_3\ra\la e^{A}_3|).
\end{equation}
The final form of the the full Hamiltonian in terms of Pauli
matrices, after neglecting additive and multiplicative constants
becomes

\begin{eqnarray}\label{HA}
    H_A&=& \sum_{i=1}^N  J_1\sigma^z_{i} \sigma^z_{i+1}+J_2\sigma^z_{i}\sigma^z_{i+2}-uJ  \sigma^x_i\sigma^x_{i+2}
    +uJ \sigma^y_i\sigma^y_{i+2}\cr &-&\frac{K}{2} \sigma^x_i
    +\frac{K}{2}\sigma^z_{i}\sigma^x_{i+1}\sigma^z_{i+2},
\end{eqnarray}
where $u=\frac{1+\cos\theta}{2}$ and
\begin{eqnarray}
  J_1 &=& J\frac{u^2-1}{2}, \\
  J_2 &=& J\frac{u^2+1}{2}-\frac{K}{2}.
  \end{eqnarray}
This is a three-body Hamiltonian with two free coupling constants.
We will say more about this in the discussion.

\section{Model B}\label{ModelB}
In accordance with our proposed notation, we denote the matrices in
this case by $B_0$ and $B_1$. In this case the matrices are the same
as in the previous case, except that one of the parameters, say $b$
is zero. There is no transformation which can put $B_1$ into
symmetric form, and we have
\begin{equation}\label{matrix1}
B_{1}=\left(\begin{array}{cc} a & 0 \\ c & d \end{array}\right).
\end{equation}
In this case no similarity transformation can put the matrix $B_1$
into symmetric form, without destroying the diagonal form of $B_0$
and hence the MPS will not be parity invariant or left-right
symmetric. From Eqs. (\ref{trace}) and (\ref{det}), we find that
\begin{equation}\label{hhh}
    1-g^2=ad \h 2\epsilon=a+d,
\end{equation}
which lead to the following final parametrization for the matrices,
where for definiteness we will show the matrices by a different
letter
\begin{equation}\label{TzSolutionB}
 B_0=\left(\begin{array}{cc} 1+g &
0 \\ 0 & 1-g \end{array}\right), \ \ \ \ B_1=\left(\begin{array}{cc}
\epsilon+g & 0 \\ c & \epsilon- g\end{array}\right).
\end{equation}
In this case again we have two free parameters in the MPS state,
namely $c$ and $g$. Moreover the matrices satisfy  the spin flip
symmetry condition (\ref{X}) with

\begin{equation}\label{XB}
    X=\frac{1+\epsilon}{2}\left(\begin{array}{cc}2g& 0 \\ c &
    1\end{array}\right)+ \frac{1-\epsilon}{2}\left(\begin{array}{cc}0& 2g \\ 1 &
    c\end{array}\right),
\end{equation}
while they do not have any symmetry under parity (i.e. there is no
matrix $\Omega$ satisfying (\ref{Omega})).\\

The eigenvalues of the transition matrix $E_B=B_0\otimes
B_0+B_1\otimes B_1$, (for the case $\epsilon=1$) now become
\begin{equation}\label{lambdaB}
    \lambda^B_1= 2(1+g)^2, \h \lambda^B_2=2(1-g)^2, \h
    \lambda^B_{3,4}=2(1-g^2),
\end{equation}
independent of the value of $c$. Figure (\ref{BFigure}), shows these
eigenvalues as a function of the parameter $g$. From this figure, a
few features can be recognized. First we note that at $g=0$, there
is a crossover between the largest and the second-largest
eigenvalues. This points to a possible MPS-quantum phase transition
at this point. Furthermore at $g=\pm 1$, there is a discontinuity in
the derivative of the second largest eigenvalue which points to a
discontinuity in the derivative of the correlation length. In view
of the general and rather broad definition of MPS quantum phase
transition, as the appearance of any discontinuity of a macroscopic
observable, these points are also points of MPS-QPT's. Moreover
since the eigenvalues do not depend on the parameter $c$, it appears
that the above points
are really lines in the space of coupling constants $c$ and $g$. \\
From (\ref{TzSolution}), it is clear that at $g=0$, the
un-normalized MPS turns into the following state

\begin{equation}\label{psiB}
    |\psi_B\ra=|\chi_+\ra^{\otimes N}+|\chi_-\ra^{\otimes N},
\end{equation}
where

\begin{equation}\label{chi}
    |\chi_{\pm}\ra=(1\pm g)|0\ra+|1\ra.
\end{equation}

\begin{figure}
\centering
    \includegraphics[width=10cm,height=8cm,angle=0]{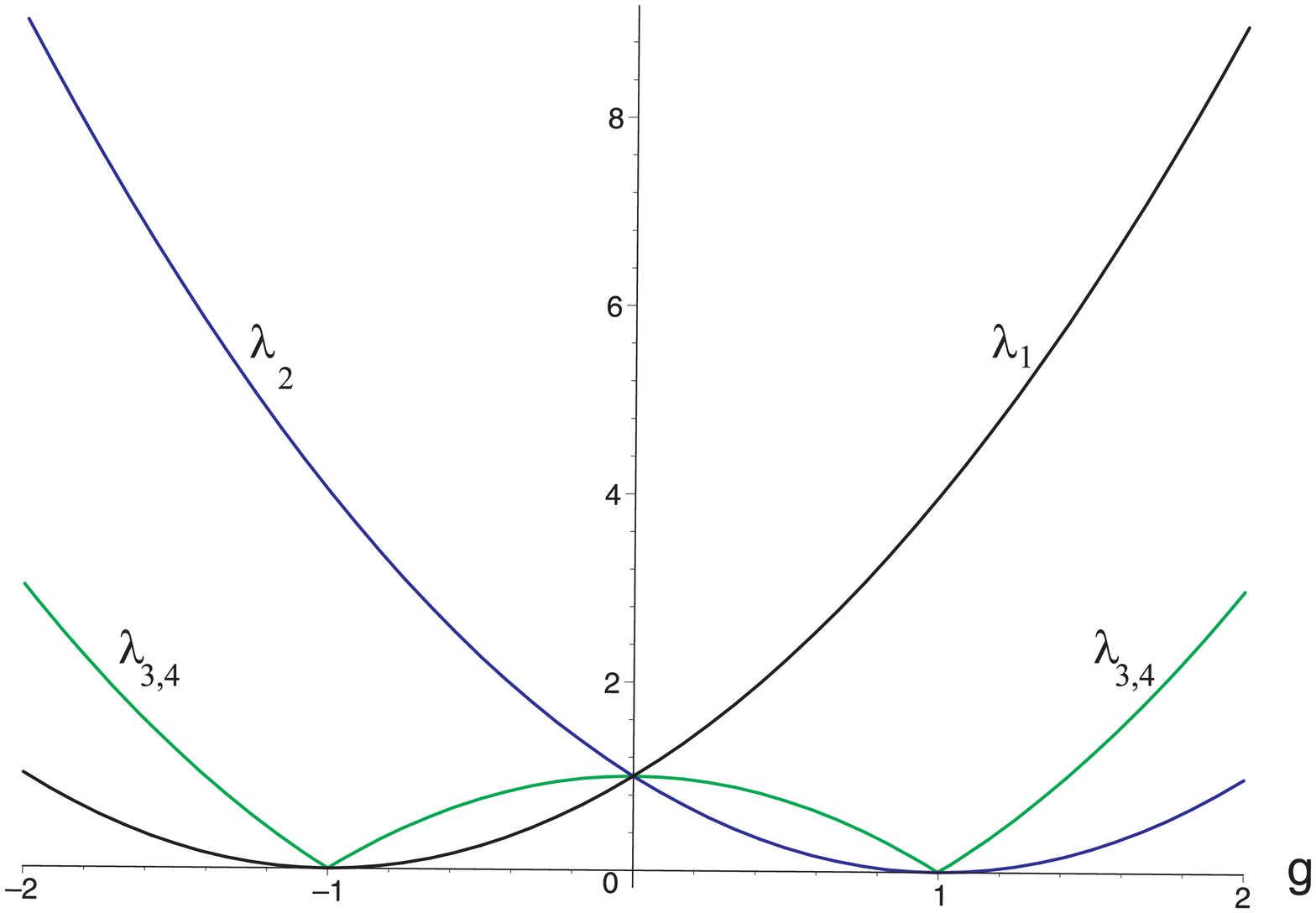}
    \caption{(Color Online) The absolute values of the eigenvalues of the transition matrix $E_B$ for $\epsilon=1$, as a function of $g$.
     }
\end{figure}\label{BFigure}

Finally we come to the parent Hamiltonian. For this model we find
that the smallest value of $k$ for which the system of equations
(\ref{cc}) has a non-trivial solution is $k=2$ and hence we can have
a two-local parent Hamiltonian. The solution space of the system of
equations (\ref{cc}) is spanned by the vectors

\begin{eqnarray}
  |e^{B}_1\ra&=&-\frac{1}{2}(1+g)|00\ra+|01\ra+\frac{1}{2}(g-1)|11\ra\\
  |e^{B}_2\ra&=&-\frac{1}{2}(1+g)|11\ra+|10\ra+\frac{1}{2}(g-1)|00\ra.
\end{eqnarray}

Interestingly we note that the above vectors transform into each
other under spin-flip, but they do not have any transformation
property under parity which is to be expected, since the original
matrices had only spin-flip symmetry. The parent Hamiltonian which
is symmetric under spin flip will be

\begin{equation}\label{hB}
    h_B=J(|e^{B}_1\ra\la e^{B}_1|+|e^{B}_2\ra\la e^{B}_2|)
\end{equation}
and the full Hamiltonian will be (after neglecting additive and
multiplicative constants and collecting all the various terms)

\begin{equation}\label{HB}
    H^B =\sum_{i=1}^N (1-g^2)
    (\sigma^x_i\sigma^{x}_{i+1}-\sigma^y_i\sigma^{y}_{i+1})+\frac{1+2g^2}{2}\sigma^z_i\sigma^z_{i+1}
    + \sigma^x_i.
\end{equation}

This is the Heisenberg XYZ system with specific couplings, that is
we have found exact solution on a submanifold in the space of
couplings. Note that since the Hamiltonian does not depend on the
paramter $c$, while the MPS does, this means that there is a large
degeneracy in the ground state. Expansion of the MPS in terms of the
parameter $c$, i.e. $|\Psi(c,g)\ra=\sum_n c^n |\psi_n(g)\ra$, will
yield the multitude of degenerate ground states $|\psi_n(g)\ra$ as
in \cite{akm}.

\section{Model C}\label{ModelC}
Denoting the matrices by $C_0$ and $C_1$, this is the only remaining
case, where  $C_0$ is not diagonalizable and can be put only in the
Jordan form
\begin{equation}\label{CJordan}
    C_0=\left(\begin{array}{cc} 1 & 0 \\ g & 1\end{array}\right).
\end{equation}
We take the general form of $C_1$ to be
$C_1=\left(\begin{array}{cc}a&b \\ c & d \end{array}\right)$ and
impose the conditions (\ref{trace}) and (\ref{det}), from which we
obtain the constraints

\begin{eqnarray}
  a+d &=& 2\epsilon \\
  ad-bc &=& 1.
\end{eqnarray}
The first constraint is solved by the parametrization $a=\epsilon+u$
and $d=\epsilon-u$, which when inserted into the second equation,
gives $u^2+bc=0$, the solution of which is $b=\mu u $ and
$c=-\frac{u}{\mu}$.  However the parameter $\mu $ can be set to
unity by a gauge transformation $C_i\lo UC_i U^{-1}$ with
$U=\left(\begin{array}{cc}1 & 0 \\ 0 & \mu\end{array}\right)$. Thus
the final form of the matrices become
\begin{equation}\label{CJordanFinal}
C_0=\left(\begin{array}{cc} 1 & 0 \\ g & 1\end{array}\right), \h
C_1=\left(\begin{array}{cc} \epsilon+u &  u \\
-u & \epsilon-u\end{array}\right).
\end{equation}
One can verify the existence of both symmetries, with matrices
\begin{equation}\label{symmC}
    X=\left(\begin{array}{cc} u & u \\ g & 0\end{array}\right)\ \ \
    {\rm and}\ \ \ \Omega = \left(\begin{array}{cc} 0 & 1 \\ 1 &
    -2\end{array}\right),
\end{equation}
such that
\begin{equation}\label{symmC2}
    X^{-1} C_i X = C_{\overline{i}},\h \Omega^{-1} C_i \Omega =
    C^T_i.
\end{equation}

The eigenvalues of the transfer matrix $E_C=C_0\otimes
C_0+C_1\otimes C_1$, are found to be

\begin{equation}\label{EigC}
\lambda^{C}_1= 2,\ \ \ \lambda^{C}_2= 2-ug,\ \ \
\lambda^{C}_{\pm}=2+\frac{ug}{2}\pm \frac{1}{2} \sqrt{16ug+u^2g^2}.
\end{equation}
Figure (\ref{CFigure}) shows the largest  eigenvalue of the transfer
matrix in the $u-g$ plane. It is seen that the lines $u=0$ and $g=0$
are the crossover lines where the largest eigenvalue changes and
hence an MPS-QPT (MPS quantum phase transition) is expected to
occur on these lines. \\

\begin{figure}
\centering
    \includegraphics[width=6.5cm,height=6cm,angle=0]{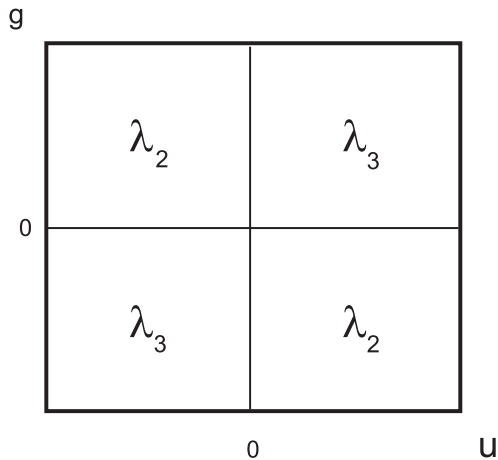}
    \caption{(Color Online) The largest eigenvalue of transfer matrix for model III. There are two crossover lines for the largest magnitude eigenvalue. }
\end{figure}\label{CFigure}

Finally we can find the parent Hamiltonian of this model. The system
of equations (\ref{cc}) has a nontrivial solution for $k=3$ and the
solution space is spanned by the vectors:

\begin{eqnarray}\label{solutionC}
    |e^{C}_1\ra&=&|001\ra+|110\ra-|011\ra-|100\ra\cr &&\cr
    |e^{C}_{2}\ra &=&
    (1+ug)(|000\ra+|111\ra)+(|001\ra+|011\ra+|100\ra+|110\ra)-3(|010\ra+|101\ra)\cr
    &&\cr
    |e^{C}_3\ra &=&
    2(|000\ra-|111\ra)-3(|001\ra-|011\ra+|100\ra-|110\ra)\cr &&\cr
    |e^{C}_4\ra &=& 2(|010\ra-|101\ra) -(1+ug)(|001\ra-|011\ra+|100\ra-|110\ra)
\end{eqnarray}
The interesting point about these vectors is that all of them are
invariant (modulo a sign) under the parity and spin-flip
transformations and hence the symmetric Hamiltonian can be written
in the following form, with four free coupling constants:

\begin{equation}\label{hC}
    h_C = \sum_{i=1}^4 J_i |e^{C}_i\ra\la e^{C}_i|.
\end{equation}

The explicit expression of the total Hamiltonian can be obtained
along the same lines as for model A and model B. We will not do it
here.

\section{Discussion}\label{Dis}
In the formalism of matrix product states, there is a large room for
constructing states and parent Hamiltonians. What really constrains
this freedom and guides us along a way which may lead to interesting
states and Hamiltonians is consideration of symmetries. The other
constraining elements is the dimension of matrices which we choose.
In this article we have classified all such states which are
constructed from two dimensional matrices and have two important
symmetries, namely the spin flip symmetry and the parity symmetry.
We have shown that there are three different models, two of which
lead to parent Hamiltonians with nearest and next-nearest
interaction (models A and C) while one of them lead to a Hamiltonian
with nearest neighbor interaction (model B). Furthermore, by
calculating the eigenvalues of the transfer matrix in each case and
determining the points of crossover between the largest and the next
to largest eigenvalues of this matrix, we have identified the points
of possible MPS quantum phase transitions. While in many of the
works which have been reported on model building in matrix product
states, \cite{mps1,mps2,mps3,mps11,mps12,mps13,mps14,mps15},
rotation symmetry has been taken into account, a condition which
highly restricts the form of matrices, in this article we have
relaxed this continuous symmetry in order to find all models
compatible with discrete symmetries in order exhaust all the
possibilities with two dimensional matrices. Any model with these
symmetries must be equivalent to one of the above three models. For
example in \cite{Cirac}, the following model was suggested

\begin{equation}\label{A'}
    A'_0=\left(\begin{array}{cc}0&0\\ 1& 1 \end{array}\right)\ ,\h A'_1=\left(\begin{array}{cc}1&q\\ 0& 0 \end{array}\right)\ ,\h
\end{equation}
which has spin-flip symmetry with $X=\left(\begin{array}{cc}0&q\\
1&0\end{array}\right)$. The Hamiltonian for this model is
\cite{Cirac}

\begin{equation}\label{ciracH}
H = \sum_{i=1}^N 2(q^2-1)\sigma^z_{i} \sigma^z_{i+1}-(1+q)^2
\sigma^x_{i}+(q-1)^2\sigma^z_{i}\sigma^x_{i+1}\sigma^z_{i+2}.
\end{equation}
It is easy to see that this model is equivalent to model A above. In
fact the transformation $$SA_i(g=1) S^{-1}=2 A'_i$$ with
$S=\left(\begin{array}{cc}0&\cos \frac{\theta}{2}\\ \sin
\frac{\theta}{2}& -\cos \frac{\theta}{2}
\end{array}\right)$
in which $q=\frac{1+\cos \theta}{2}$, proves the equivalence.

\section{Acknowledgements} I would like to thank V.
Karimipour for valuable discussion and comments.

{}

\end{document}